\documentclass{article}
\usepackage{graphicx} 
\usepackage{amsmath, amssymb, amsfonts}
\usepackage{tikz}
\usepackage{appendix}
\usepackage{listings}
\usepackage{color}
\usepackage{pgfplots}
\usepackage{authblk}
\usepackage{xcolor}
\usepackage[
backend=biber,
style=alphabetic,
sorting=ynt
]{biblatex}
\usepackage[
frozencache=true,
cachedir=minted-cache
]{minted} 

\definecolor{LightGray}{gray}{0.95}

\tikzset{>=latex}
\newcommand{\reals}[0]{\mathbf{R}}
\newcommand{\ones}[0]{\mathbf{1}}
\newcommand{\beq}[0]{\begin{equation}}
\newcommand{\eeq}[0]{\end{equation}\vspace{1pt}}
\newcommand{\baq}[0]{\begin{align}}
\newcommand{\eaq}[0]{\end{align}\vspace{1pt}}
\newcommand{\ie}[0]{\textit{i.e.}}
\newcommand{\eg}[0]{\textit{e.g.}}

\newcommand{\redline}[0]{\raisebox{2pt}{\tikz{\draw[thick, red](0,0) -- (5mm,0)}}}
\newcommand{\blueline}[0]{\raisebox{2pt}{\tikz{\draw[thick, blue](0,0) -- (5mm,0)}}}

\title{A Tick-by-Tick Solution for Concentrated Liquidity Provisioning}
\author{Corinne Powers\thanks{corinne@gadget.fund}}
\affil{Gadget Capital}
\date{\today}

\begin{document}

\maketitle

\begin{abstract}
Automated market makers with concentrated liquidity capabilities are programmable at the tick level. The maximization of earned fees, plus depreciated reserves, is a convex optimization problem whose vector solution gives the best provision of liquidity at each tick under a given set of parameter estimates for swap volume and price volatility. Surprisingly, early results show that concentrating liquidity around the current price is usually not the best strategy.
\end{abstract}

\section{Introduction}
Consider the setting where capital is provisioned via an automated market maker (AMM) across a set of liquidity pools, \ie, ticks, where each tick earns fees in exchange for acting as counter-party for swaps that occur within a specific price range. Each tick may already have liquidity provisioned to it, and fees are distributed pro rata amongst liquidity providers. 

To provision liquidity to a tick, a \textit{liquidity provider} must add a specific combination of tokens depending on the current state of the AMM. (While it's possible for a liquidity pool to support swaps between baskets of two or more tokens \cite{basket}, we consider an asset-stablecoin pair.) 

For the purpose of tick-by-tick provisioning, this usually takes the form of ``100\% asset'' or ``100\% stablecoin'' depending on whether the tick price range is above or below the current price, respectively, with an asset-stablecoin combination at the current price. As the asset price moves, the combination and value of a tick's reserves also changes.

In general, it seems clear that provisioning liquidity to ticks near the current price is expected to earn more swap fees. As price bounces up and down, swaps are pushed into and through neighboring ticks and contribute to earned fees.  In fact the more concentrated the provisioning within a tick, the larger the allocation of earned fees.

On the other hand, as prices drop a tick's reserves depreciate, making it more expensive to re-balance. This risk is higher for positions that are more concentrated. Intuitively, there is a balance to strike between earning fees quickly under concentration and keeping ticks active longer under dispersion.

There are other contributing factors as well. For example, ticks already saturated with liquidity will have lower return on investment for a fixed amount of trade volume, due to the pro rata fee distribution. And one may wish to incorporate their predictions about price volatility and swap volume to focus liquidity where they think it will return the best yield. 

In this work, we look at two ways to provision liquidity. The first aims to capture the most fees using a portfolio of single-tick ranges, introducing the reader to tick-by-tick provisioning and solving for the exact solution. Next we study the more relevant problem of maximizing returns (net of earned fees and depreciated reserves) which we show is a convex optimization problem. 

We've framed the problem to accommodate ticks drawn from any protocol, fee tier or network, which are consolidated into a ``candidate tick set'' that are considered for provision.

\section{Background}
There is sizable literature in recent years that focus on automated market makers. A common thread is their relationship to limit orders books, call and put options \cite{lambert2023panoptic}, and the performance gap lost to arbitrageurs \cite{milionis2023automated} using tools from Black-Scholes options pricing \cite{blackscholes}.

Tools from convex analysis have been used to study the general case of constant function market makers, as shown in \cite{Angeris_2020} and \cite{angeris2021replicating}. And to show how automated market makers can be extended to handle many types of swaps \cite{basket}.

The interested reader may turn to the ``Related literature'' section of \cite{milionis2023automated} for a thorough background on these topics.

We direct our work to an audience of liquidity providers, for whom earned fees is an important part of their objective.

\section{Provisioning Problems}
In deriving our approach to tick-by-tick provisions, we make the following assumptions:
\begin{itemize}
\item Provisions remain in place for a fixed period of time,
\item No liquidity is added or removed by any provider during this period,
\item There are no transaction costs to open, close, or rebalance provisions,
\item There exist largely deep order books and alternative markets, obviating the risk of price manipulation.
\end{itemize}

Under these assumptions, the value of a portfolio of tick-by-tick provisions after a fixed period of time will depend only on the AMM state, trade volume, and latest price.

\paragraph{Maximum revenue.}
Consider the optimization problem,

\beq\label{eq:maxrev}
\begin{array}{ll}
    \mbox{maximize} & \displaystyle\sum_{i=1}^n a_i\frac{x_i}{x_i + b_i} \\~\\
    \mbox{subject to} & x \ge 0, \qquad \ones^Tx = d,
\end{array}
\eeq

where $a_i, b_i \ge 0$ for $i = 1, \ldots, n$ and $d \ge 0$. This is the \textit{maximum revenue} liquidity provisioning problem for AMMs with pro rata distributions and tick-level pools. For each tick $i$, parameter $a_i$ is estimated revenue earned from fees over a fixed time period, and $b_i$ is current liquidity. The parameter $d$ is the total value to allocate across all ticks. The variable $x_i$ is the value of reserves provisioned to tick $i$, and $x_i/(x_i + b_i)$ gives the fraction of remitted fees (revenue) granted under that provision. 



\paragraph{Solution method via water-filling.}
The solution to problem (\ref{eq:maxrev}) is
\beq
x_i = \max\left\{0, \sqrt{a_ib_i}(u - \sqrt{b_i/a_i})\right\}, \qquad i = 1, \ldots, n, 
\eeq
where $u$ is determined by the condition
\beq\label{eq:water-fill}
\displaystyle\sum_{i=1}^n \max\left\{0, \sqrt{a_ib_i}(u-\sqrt{b_i/a_i})\right\} = d.
\eeq
(The proof for this can be seen in Appendix \ref{sec:water-fill-proof}.)
The left hand side of equation (\ref{eq:water-fill}) is a piece-wise linear increasing function of $u$, with breakpoints at $\sqrt{b_i/a_i}$. So the equation has a unique solution which is readily determined.

\begin{figure}

\begin{center}
\begin{tikzpicture}

\fill[gray!30!white] (0,0) rectangle (8.00, 2.50);
\filldraw[fill=white, draw=black] (8.00, 0) -- (0,0) -- (0, 2.21) -- (0.37, 2.21) -- (0.37, 0.79) -- (1.45, 0.79) -- (1.45, 4.40) -- (1.80, 4.40) -- (1.80, 1.37) -- (2.94, 1.37) -- (2.94, 2.07) -- (3.60, 2.07) -- (3.60, 1.48) -- (4.40, 1.48) -- (4.40, 3.14) -- (5.37, 3.14) -- (5.37, 2.90) -- (6.07, 2.90) -- (6.07, 3.00) -- (6.52, 3.00) -- (6.52, 1.33) -- (6.85, 1.33) -- (6.85, 1.42) -- (7.06, 1.42) -- (7.06, 1.27) -- (8.00, 1.27);

\draw (0,0) -- (0,5);
\draw (0,0) -- (8,0);
\node[draw=white, anchor=east] at (-0.1, 2.5) {$u$};
\node[draw=white, anchor=north] at (4.0, -0.1) {$i$};

\draw[<->] (8.20, 0) -- (8.20, 1.27);
\node[draw=white, anchor=west] at (8.30, 0.63) {$\sqrt{b_i/a_i}$};
\draw[<->] (7.06, -0.2) -- (8.00, -0.2);
\node[draw=white, anchor=north] at (7.53, -0.3) {$\sqrt{a_ib_i}$};
\draw[dashed, thin] (7.06, 1.27) -- (7.06, 2.50) -- (8.00, 2.50) -- (8.00, 1.27);
\node[draw=gray!30!white] at (7.53, 1.9) {$x_i$};

\end{tikzpicture}
\caption{Illustration of water-filling algorithm for Problem (\ref{eq:maxrev}). The height of each region is $\sqrt{b_i/a_i}$ and the width is $\sqrt{a_ib_i}$. The region is flooded to water level $u$ which uses a total quantity of water equal to $d$. The amount of water above each patch is the optimal value $x_i$.}
\label{fig:water-fill}
\end{center}
\end{figure}
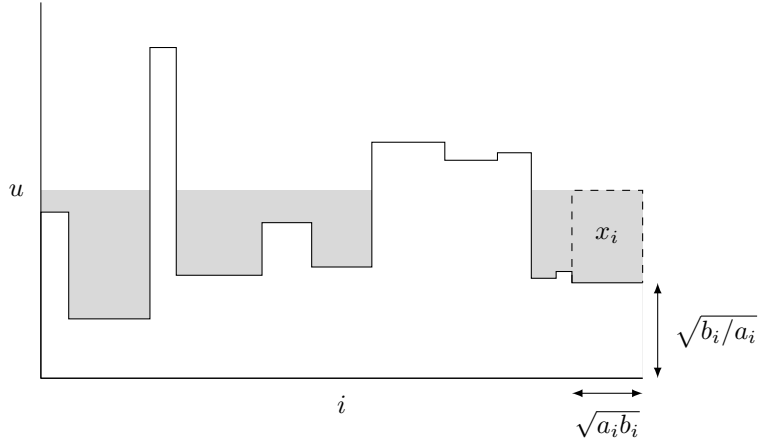

This solution method is called \textit{water-filling} for the following reason. We think of mounds of earth on patch $i$ with height $\sqrt{b_i/a_i}$ and width $\sqrt{a_ib_i}$ that form a landscape, and flood the region with $d$ total water so that its surface height is $u$ as shown in Figure (\ref{fig:water-fill}). The volume of water above patch $i$ is then the optimal value $x_i$.

Ticks with large predicted revenue, $a_i$, will see proportionately more water allocated at a faster rate as the surface level rises. Meanwhile, a large amount of existing liquidity, $b_i$, is a hurdle to clear before accepting a smaller rate of return on investment, \ie, a worse level of diminishing return.

\paragraph{Maximum return.}
Now consider an variation of Problem (\ref{eq:maxrev}) that includes an additional term,
\beq\label{eq:maxret}
\begin{array}{ll}
    \mbox{maximize} & \displaystyle\sum_{i=1}^n \left(a_i\frac{x_i}{x_i + b_i} + c_ix_i\right) \\~\\
    \mbox{subject to} & x \ge 0, \qquad \ones^Tx = d,
\end{array}
\eeq
where $c_i \ge 0$ for $i = 1, \ldots, n$. This is the \textit{maximum return} liquidity provisioning problem, where $c_i$ is a scaling factor for tick $i$'s return of reserves in expectation over price, \ie, $c_i$ is the expected value of tick $i$'s reserves one time period after provisioning one unit of capital (\eg, \$1). 
If $c_i < 1$ ($c_i > 1$) then tick $i$'s reserves are expected to depreciate (appreciate), and the sum, $\sum_i c_i x_i$, is the expected total value of reserves across all ticks.

We refer to $(a, b, c)$ as the market conditions, $d$ as the total capital, and $x$ as the optimal provision.

\paragraph{Solution method via convex optimization.}
Problem (\ref{eq:maxret}) is a convex optimization problem \cite{boyd2004} because the objective is concave maximization and the constraint set is a (scaled) probability simplex. Using the identity $-x/(x + b) = b/(x + b) - 1$, Problem (\ref{eq:maxret}) is equivalent to
\beq \label{eq:water-fill-min}
\begin{array}{ll}
    \mbox{minimize} & \displaystyle\sum_{i=1}^n \left(\frac{a_ib_i}{x_i + b_i} - c_ix_i - a_i\right) \\~\\
    \mbox{subject to} & -x \preceq 0, \qquad \ones^Tx = d,
\end{array}
\eeq
which is the standard form of a convex problem. 

Therefore a global solution can be certified to a given level of accuracy. Also, software tools and computational solvers are widely available and relatively easy to use in practice. It is possible for liquidity providers to iterate quickly using this approach under changing market conditions.

Figure (\ref{code:maxret}) walks through the steps of solving Problem (\ref{eq:maxret}) using \texttt{cvxpy} \cite{diamond2016cvxpy}, a Python embedded modeling language for convex optimization problems. The software performs best when parameters $a$, $b$, $c$, and $d$ are scaled to have values near 1, rather than, say, values near 1 million. We offer tips and approaches for estimating these parameters from historical data and market intelligence in Section {\ref{sec:params}}.

\begin{figure}[htb!]
\centering
\begin{minipage}{0.8\textwidth}
\begin{minted}{python}
import cvxpy as cp

# specify parameters and variables
a = cp.Parameter(n, pos=True)
b = cp.Parameter(n, pos=True)
c = cp.Parameter(n, pos=True)
d = cp.Parameter(pos=True)
x = cp.Variable(n, pos=True)

# set parameter values
a.value = fee_by_tick
b.value = liquidity_by_tick
c.value = reserves_return_by_tick
d.value = total_capital

# construct the convex optimization problem
obj = cp.Minimize(
    cp.multiply(a, b)@cp.inv_pos(x+b)
    - c@x - cp.sum(a)
) 
constraints = [cp.sum(x) == d, x >= 0]
prob = cp.Problem(obj, constraints)

# solve
best_return = prob.solve()
best_provision = x.value
\end{minted}
\end{minipage}
\caption{A code snippet from a solution to the maximum return provisioning problem implemented using the \texttt{cvxpy} software package.
}
\label{code:maxret}
\end{figure}

\section{Parameter Estimation}\label{sec:params}

\paragraph{Current liquidity by tick, $b$, is known.}
The amount of liquidity provisioned to the current tick usually appears as a state of the AMM. 

Additional processing may be required to transform net liquidity changes from neighboring ticks into an absolute value, but the result is typically easy to verify using online tools such as `info.uniswap.org' for Uniswap V3 ticks.

\paragraph{Use swap volume to predict $a$, future fees by tick.} 
As evident from the water-filling solution to Problem (\ref{eq:maxrev}), swap volume, as well as price volatility, plays a significant role in the optimal provision. There are two important aspects to swap volume: shape and size.

One approach to swap volume estimation uses signals processing techniques for non-parametric fitting. 
For example, denoise a (centered) $\reals^n$-valued signal over $T$ time periods in order to uncover a latent (possibly dynamic) market signal, 
whose noisy estimates are observed at every time period. 

Another approach is to fit statistical models, either jointly over shape and size, or separately. For example, fit a Gaussian shape to centered and scaled tick-wise swap volumes taken from a range of time periods. Then scale by total swap volume averaged over multiple time periods.

One may incorporate market intelligence into estimates. If an upcoming event is expected to drive higher trade volume (not only higher volatility), then estimates derived from historical data may be manually adjusted to reflect the anticipation.

There is a particular type of data processing step that is frequently needed for proper calculation of historical swap volume by tick. Swaps that start in tick $i$ and end in tick $j$ generate swap volume that appears in tick $i$, tick $j$, and all ticks in between. It is possible to determine exactly how much volume is swapped by each tick using details about the AMM, such pre- and post-swap price, as well as liquidity for every tick along the price path.

The final step is to scale tick-wise swap volume by the tick's fee tier, which is known. Fee tiers are usually 1, 5, 30, or 100 basis points, passing that fraction of total swap volume on to providers as fees earned for their service.

We may consider candidate ticks from pools with different fee tiers, \eg, ticks associated with \texttt{ETH-USDC} swaps for pools with 0.05\% and 0.3\% fee tiers. For each fee tier, swap volume by tick will be different, and of course the fee multiplier will be different.

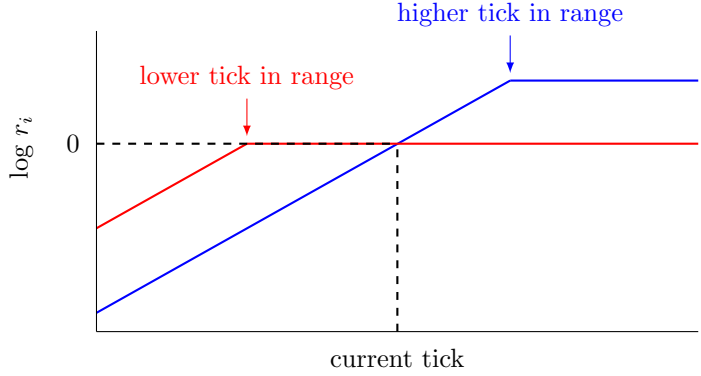
\begin{figure}[htb!]
\begin{center}
\begin{tikzpicture}
\draw[blue, thick] (0, 0.25) -- (5.5, 3.34);
\draw[blue, thick] (5.5, 3.34) -- (8, 3.34);
\draw[red, thick] (0, 1.375) -- (2, 2.5);
\draw[red, thick] (2, 2.5) -- (8, 2.5);
\draw[thick, dashed] (4, 0) -- (4, 2.5);
\draw[thick, dashed] (0,2.5) -- (4, 2.5);
\draw (0,0) -- (0,4);
\draw (0,0) -- (8,0);
\node[draw=white, anchor=east] at (-0.1, 2.5) {0};
\node[draw=white, anchor=north, text width=1.8cm, align=center] at (4.0, -0.1) {current tick};
\draw[<-, blue] (5.5, 3.44) -- (5.5, 3.94) node[anchor=south] {higher tick in range};
\draw[<-, red] (2, 2.6) -- (2, 3.1) node[anchor=south] {lower tick in range};
\node[anchor=east, rotate=90] at (-1, 3) {log $r_i$};
\end{tikzpicture}
\end{center}
\caption{Log return of reserves for two ticks from a Uniswap V3 pool. One tick is active at a higher price (\protect\blueline) while the other is active at a lower price (\protect\redline).}
\label{fig:reserves-tick}
\end{figure}

\paragraph{Use price volatility to predict $c$, expected return of reserves by tick.}
For any tick, its return of reserves, $r_i \in \reals_+^n$, is a deterministic function of asset price and invariant under time and price path. Consider the two examples of Uniswap V3 ticks illustrated in Figure (\ref{fig:reserves-tick}), which are displayed on a log scale for clarity. 

Reserves provisioned to a ``higher'' tick start out as only asset, and increase in value while the price rises until the tick is in range and asset is swapped for stable. Meanwhile reserves provisioned to a ``lower'' tick start out as only stable, and maintain constant value while the price drops until the tick is in range and stable is swapped for asset. For both, asset value shrinks as price drops, and hits a ceiling as price rises. 

Notice that the expected return, $c_i$, for tick $i$ depends on next period price, which is a random variable with mass function $p \in \reals_+^n$. So
\begin{equation}
c_i =p^T r_i, \qquad i = 1, \ldots, n.
\end{equation} 
And so the sum $\sum_{i=1}^n c_i = p^T\left(\sum_{i=1}^n r_i\right)$ is expected value of a portfolio after one time period if each tick is provisioned one unit of capital (\eg, \$1).

By definition, volatility is the standard deviation of $p$, the next-period price, multiplied by the square root of the number of time periods in, say, a year. A common model for price is geometric Brownian motion, \ie, one assumes log price returns are iid Gaussian so that the model is straight forward to fit using historical data.

Finally, it is possible to consider a candidate pool of ticks from protocols with different swap criteria (e.g., constant product- and constant sum- market makers) because $r_i$ may be tailored separately for each tick $i$.

\paragraph{Models for volatility and volume should be consistent.}
It's possible for swap volume by tick to have a different shape than the distribution of prices after one period. This may be due to a number of factors, such mid-period events, concentrated liquidity, order book depth, and more. 

Yet there are some swap volume curves (\eg, low entropy concentrated volume) that are inconsistent with price change curves (\eg, high entropy price dispersion) even under highly manipulative conditions. Or swap volume may be centered around a price that's different from the center of future prices. Solutions to Problem (\ref{eq:maxret}) where parameters $a$ and $c$ are generated from inconsistent volatility and volume models may look strange, even though the objective is properly optimized.

\section{Experiments}
We begin with a qualitative test for the purpose of exploration. Consider the setting where $d$ capital may be allocated across ticks from a candidate set, $\Omega$. The current price is $p$ and the provisions will be held for a time period of $T$ days. To estimate $a$, determine the standard deviation of swap volume data from the previous $T$ days, and center a Gaussian with that standard deviation at the current price. To estimate $c$, use implied volatility, $\eta$, to generate of distribution of where prices will be after $T$ days. 

In our experiment, $d$ is \$1 million, $T$ is 7 days, and $\Omega$ is the set of ticks for the \texttt{ETH-USDC} pair with 0.05\% fee from Uniswap V3 on Ethereum. The price $p$ was \$2780 on February 16, 2024, and so ticks were limited further still to the range of \$2500 to \$3060, or $\pm10\%$ of the current price. We choose implied volatility $\eta = 80\%$ because it appears to be roughly the 7-day moving average of implied volatility estimates from options data across various trading platforms. We used the `web3.py' software package to pull liquidity, price, and swap volume data directly from the Uniswap V3 protocol. 

\begin{figure}[htb!]

    \centering
    \includegraphics[width=5in]{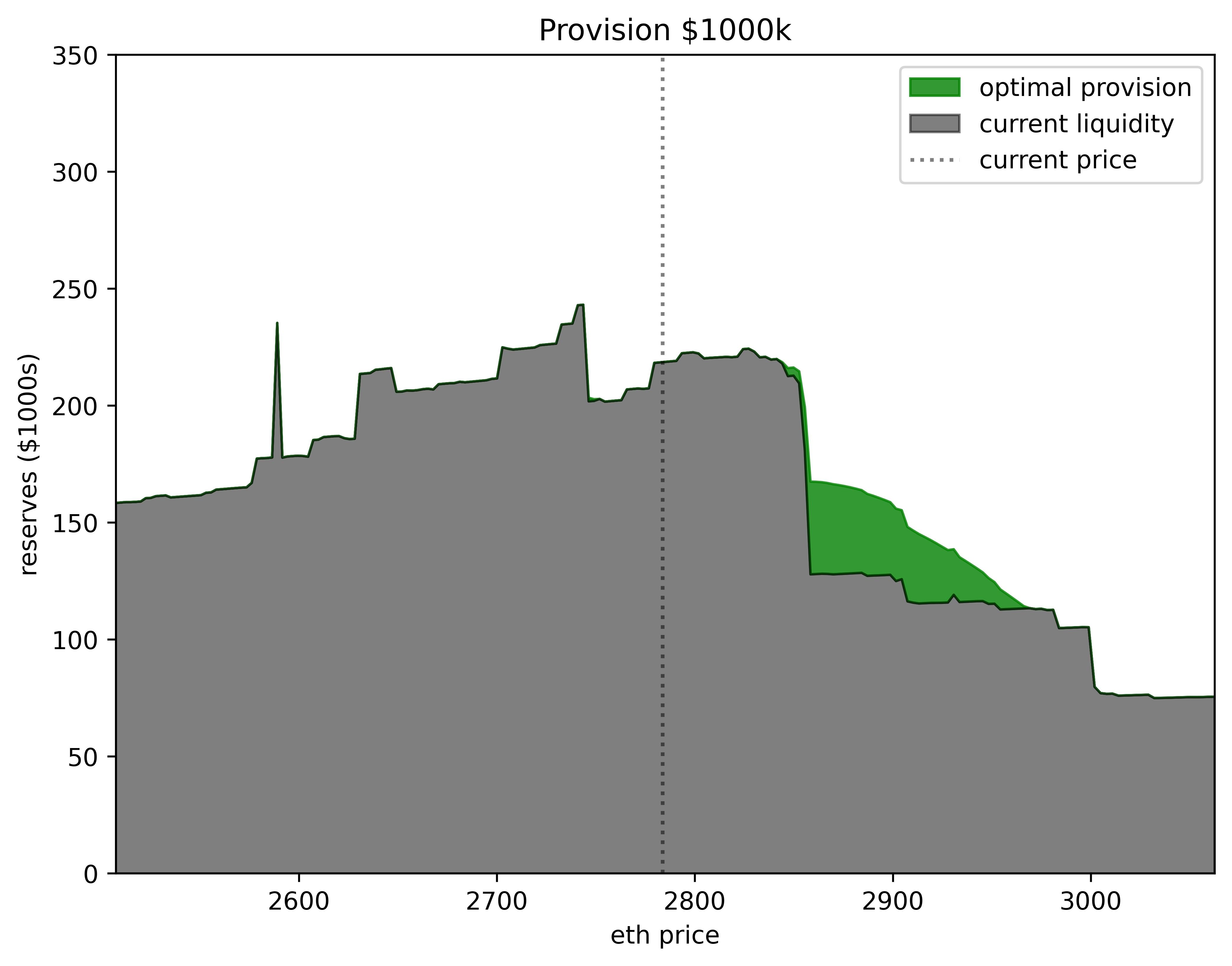}
    \caption{Given current liquidity, the optimal tick-by-tick provision is concentrated in the well closest to the current asset price.}
    \label{fig:results}
\end{figure}

Figure (\ref{fig:results}) illustrates liquidity, $b$, already in the pool as well as provisions, $x$, in green. First, notice that there are no provisions at the current tick. Rather, almost all liquidity is provisioned to ticks in the \$2830 to \$2980 price range. One reason for this is there is a drop off in liquidity at that range, implying a larger portion of fees for the same amount of capital. Second, we notice that the amount provisioned to each tick tapers off as the price gets larger. Presumably the rate of tapering depends on the trade-off between price volume and volatility. 

\begin{figure}[htb!]

    \centering
    \includegraphics[width=\linewidth]{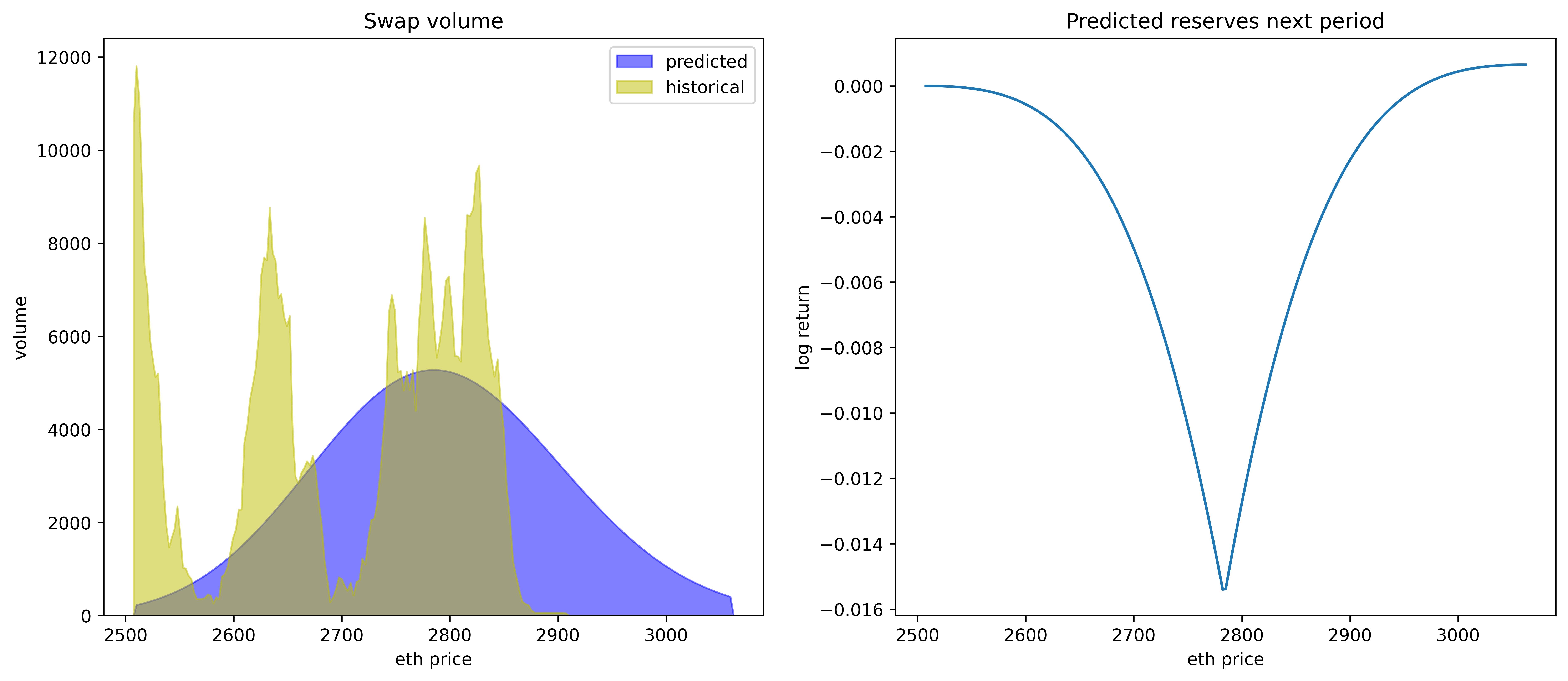}
    \caption{On the left, historical swap volume (yellow) and predicted swap volume (purple) share the same standard deviation and total size but take different shapes. On the right, reserves provisioned to the current tick have the most value to lose.}
    \label{fig:estimates}
\end{figure}

We consider these two effects in Figure (\ref{fig:estimates}). On the left, previous swap volume is shown in yellow. Even though the yellow historical data not appear Gaussian in shape, the purple distribution centered at the current price is our estimate of future volume, which is proportional to $a$. To the right is the expected return on reserves, $c$, under the assumption that asset prices following geometric Brownian motion with implied volatility determined by the market. Notice that the greatest loss occurs at the current price. Among ticks with the most at stake (\ie, ticks at, or above, current price) it also has the least to gain.

\begin{figure}[htb!]

    \centering
    \includegraphics[width=\linewidth]{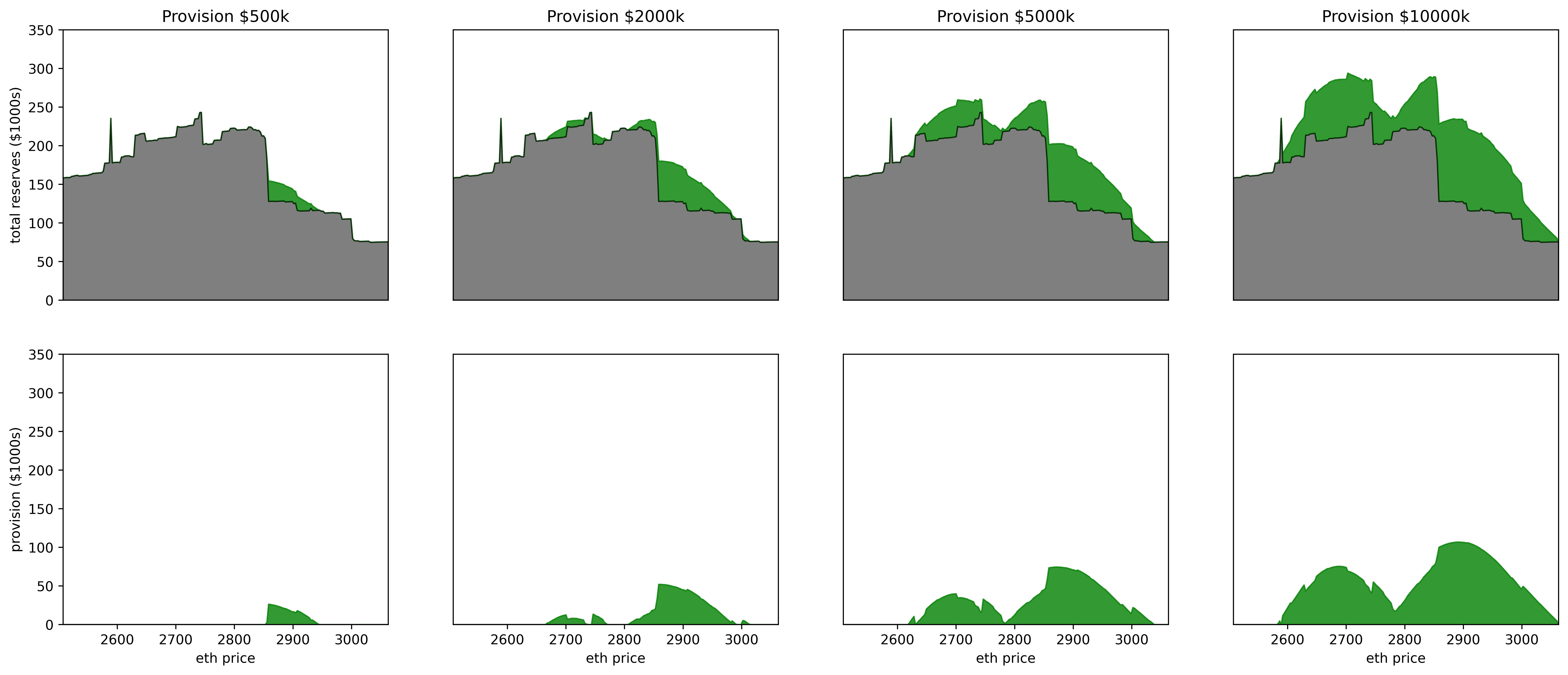}
    \caption{From left to right, the total capital available to provision across all ticks increases from \$200k to \$10m. Under optimal provisions, the total liquidity now available in the pool is shown above while the tick-by-tick provisions are isolated below.}
    \label{fig:drange}
\end{figure}

Now consider the setting where we may choose how much total capital, $d$, to provision. Figure (\ref{fig:drange}) illustrates the shape that tick-by-tick provisions follow as more capital is allocated to the pools. The top row illustrates the overlay of new provisions to current liquidity, and the bottom row shows only the tick-by-tick provisions as determined by Problem (\ref{eq:maxret}). From the top, it's clear that provisions continue to seek a larger portion of fees by piling more quickly into wells. We call this the `fee-chasing' effect. We also notice a bi-modal shape emerging as $d$ increases. We say this is due to the reserves `loss-avoiding' effect, which scales linearly in $d$ and thus becomes more prominent at large values. To our knowledge, this phenomena is new to the community of liquidity providers. 

Finally, we wish to understand how the performance of tick-by-tick provisioning compares to that of single-range provisioning, where provision are equally distributed across ticks and centered around the current price. For good measure, we consider the delta-neutral version as well, where a short position equal to the value of reserves across all assets in ticks is opened to remove asset exposure at the current price.

Consider the setting where the tick-by-tick model is fit to data from the previous $S$ days. Then market behavior is simulated for $T$ days before the positions are closed. The new value of reserves and earned fees determines performance. This process is repeated every $R$ days.

Table (\ref{tab:perf}) shows the results of this experiment for a single range tick spanning ticks at $\pm 10\%$ of the current price, where $T = S = 7$ days and $R = 4$ days. (We use block numbers to identify the beginning of train and test periods.) This experiment roughly covers data from January 1, 2024 -- February 16, 2024, and corresponds to a period of roughly 30\% growth in the price of \texttt{ETH}. 

The results show that, for this set of market conditions, tick-by-tick provisioning performed better in expectation but performance fluctuated more each week. The single range strategy benefited from upside exposure in the underlying asset, which hurt the performance of the delta-neutral approach. The authors conclude that further testing is needed to compare the performance under different market conditions. Meanwhile, the results appear to be promising.

\begin{table}[htb!]
\label{tab:perf}
\begin{center}
\label{tab:benchmark}
\begin{tabular}{|cc|c|c|c|}\hline
Train block & Test block & Tick by tick & Range $\pm 10\%$ & Delta-neutral \\ \hline \hline
18943206 & 18993312 & 4.24\% & 4.50\% & 0.49\% \\
18971817 & 19021923 & 4.00\% & 4.48\%&  -1.11\% \\
19000435 & 19050541 & 1.69\% & 3.86\% & -2.57\% \\
19029031 & 19079137 & 10.63\% & 5.25\% & -8.56\% \\
19057636 & 19107742 & 8.67\% & 5.30\% & -6.25\% \\
19086319 & 19136425 & 7.38\% & 4.96\% & -6.66\% \\
19114856 & 19164962 & 3.55\% & 4.79\% & -5.97\% \\
19143467 & 19193573 & 1.1\% & 3.80\% & -3.71\%\\ \hline\hline
\multicolumn{2}{|r|}{mean} & 5.16\% & 4.62\% & -4.29\%\\
\multicolumn{2}{|r|}{std} & 3.17\% & 0.54\% & 2.89\% \\ \hline
\end{tabular}
\caption{Compare the results of tick-by-tick provisions to a $\pm 10\%$ single-range position, also with a delta neutral hedge. Deploy and rebalance \$1 million on a weekly basis from Jan 1, 2024 -- Feb 16, 2024.}
\end{center}
\end{table}

\newpage
\section{Discussion}
Tick-by-tick provisioning offers a different way to think about programming liquidity to the decentralized economy. There are competing effects that are unique to liquidity providers, and we use vector and convex math, rather than financial greeks (\eg, delta, theta, gamma), to strike this balance without leaving money on the table. We also separate out the effects due to price volume from those due to volatility. In our experience, these two effects are often conflated.  

There remain many open questions and the need for further results in order to determine how best to utilize the framework. To name just a few, 
\begin{itemize}
\item What does a multi-pool tick provision look like? In what capacity are multi-fee provisions optimal and are there any fee tiers missing? 
\item How sensitive are the provision results to changes in estimate of volatility and volume? What more can we say about the empirical relationship between volatility and volume?
\item How does just-in-time liquidity (JIT) fit in? Rather than initialize a far away tick now, can we provision the liquidity when the price moves closer? How close? If everyone's doing this, what does it mean for our current liquidity parameter, $b$?
\item Is this approach delta neutral? If we look at the change in return with respect to the change in price, what shape does it take?
\item How else can we interpret the avoidance around current price? Is there a connection to LVR \cite{milionis2023automated}?
\item What else can we do when Uniswap V4 is released?
\end{itemize}

Most of all, we're excited by the prospect of the wider implications of these ideas. Could it be possible that optimal provisioning is not actually centered around the current price? If so, what would that mean for the community? Could this drive deeper liquidity to decentralized markets?

\appendix
\section{Appendix}
\subsection{Water filling method for maximum revenue solution}\label{sec:water-fill-proof}
(The following proof is a direct extension of an example given in \cite{boyd2004} on page 245, and so the wording may be similar.)

Recall the maximum revenue problem (\ref{eq:maxrev}). Using the fact that $x/(x + b) = 1 - b/(x + b)$, the objective can be rewritten as
\beq
\displaystyle\sum_{i=1}^n\frac{a_ib_i}{x_i + b_i}.
\eeq
\\
Introduce Lagrange multipliers $\lambda_i \in \reals^n$ for the inequality constraints $x \ge 0$, and a multiplier $\mu \in \reals$ for the equality constraint $\ones^Tx = d$. Then the KKT conditions are
\beq
\begin{array}{c}
x \ge 0, \qquad \ones^Tx = d, \qquad \lambda \ge 0,\\~\\
x_i\lambda_i = 0, \qquad i = 1, \ldots, n,\\~\\
-\frac{a_ib_i}{(x_i+b_i)^2} - \lambda_i + \nu = 0, \qquad i = 1, \ldots, n.
\end{array}  
\eeq
Note that $\lambda_i$ is a slack variable and can be eliminated, leaving
\beq
\begin{array}{c}
x \ge 0, \qquad \ones^Tx = d, \qquad x_i\left(\nu -\frac{a_ib_i}{(x_i+b_i)^2} \right) = 0, \qquad i = 1, \ldots, n,\\~\\
\frac{a_ib_i}{(x_i+b_i)^2} \le \nu, \qquad i = 1, \ldots, n.
\end{array}  
\eeq
Focus on the last inequality. Invert the terms and take the square root,
\beq
\frac{x_i+b_i}{\sqrt{a_ib_i}} \ge \frac{1}{\sqrt{\nu}}, \qquad i = 1, \ldots, n, 
\eeq
then substitute $u = \frac{1}{\sqrt{\nu}}$ to get
\beq
\frac{x_i+b_i}{\sqrt{a_ib_i}} \ge u, \qquad i = 1, \ldots, n.
\eeq
The new KKT conditions are
\beq
\begin{array}{c}
x \ge 0, \qquad \ones^Tx = d, \qquad x_i\left(u - \frac{x_i + b_i}{\sqrt{a_ib_i}} \right) = 0, \qquad i = 1, \ldots, n,\\~\\
\frac{x_i+b_i}{\sqrt{a_ib_i}} \ge u, \qquad i = 1, \ldots, n.
\end{array}  
\eeq
We can directly solve these equations to find $x_i$ and $u$. If $u > \sqrt{b_i/a_i}$, the last condition can only hold if $x_i > 0$, which by the third condition implies that $x = \sqrt{a_ib_i}(u - b_i)$. If $u \le \sqrt{b_i/a_i}$, then $x_i > 0$ is impossible, because it would imply $u \le \sqrt{b_i/a_i} < \frac{x_i+b_i}{\sqrt{a_ib_i}}$, which violates the complementary slackness condition. Therefore, $x_i = 0$ if $u \le \sqrt{b_i/a_i}$. Thus we have
\beq
x_i = \left\{
\begin{array}{ll}
\sqrt{a_ib_i}(u - \sqrt{b_i/a_i}) & u > \sqrt{b_i/a_i}, \\
0 & u \le \sqrt{b_i/a_i}.
\end{array}
\right.
\eeq
or, more simply put, $x = \max\left\{0, \sqrt{a_ib_i}(u - \sqrt{b_i/a_i})\right\}$. Substituting this expression into the condition $\ones^Tx = 1$ we obtain the water-filling condition,
\beq
\displaystyle\sum_{i=1}^n \max\left\{0, \sqrt{a_ib_i}(u - \sqrt{b_i/a_i})\right\} = d. 
\eeq

\printbibliography

@article{diamond2016cvxpy,
  author  = {Steven Diamond and Stephen Boyd},
  title   = {{CVXPY}: {A} {P}ython-embedded modeling language for convex optimization},
  journal = {Journal of Machine Learning Research},
  year    = {2016},
  volume  = {17},
  number  = {83},
  pages   = {1--5},
}

@book{boyd2004,
  author  = {Stephen Boyd and Lieven Vandenberghe},
  title   = {Convex Optimization},
  year    = {2004},
  publisher = {Cambridge University Press},
}

@misc{basket,
url = {https://arxiv.org/pdf/2107.12484.pdf},
    doi = {2107.12484},
    author = {Guillermo Angeris and Akshay Agrawal and Alex Evans and Tarun Chitra and Stephen Boyd},
    title = {Constant Function Market Makers: Multi-Asset Trades via Convex Optimization},
    year = {2021}
}

@misc{lambert2023panoptic,
      title={Panoptic: the perpetual, oracle-free options protocol}, 
      author={Guillaume Lambert and Jesper Kristensen},
      year={2023},
      eprint={2204.14232},
      archivePrefix={arXiv},
      primaryClass={q-fin.PR}
}

@misc{milionis2023automated,
      title={Automated Market Making and Loss-Versus-Rebalancing}, 
      author={Jason Milionis and Ciamac C. Moallemi and Tim Roughgarden and Anthony Lee Zhang},
      year={2023},
      eprint={2208.06046},
      archivePrefix={arXiv},
      primaryClass={q-fin.MF}
}

@inproceedings{Angeris_2020, series={AFT ’20},
   title={Improved Price Oracles: Constant Function Market Makers},
   url={http://dx.doi.org/10.1145/3419614.3423251},
   DOI={10.1145/3419614.3423251},
   booktitle={Proceedings of the 2nd ACM Conference on Advances in Financial Technologies},
   publisher={ACM},
   author={Angeris, Guillermo and Chitra, Tarun},
   year={2020},
   month=oct, collection={AFT ’20} }

@misc{angeris2021replicating,
      title={Replicating Market Makers}, 
      author={Guillermo Angeris and Alex Evans and Tarun Chitra},
      year={2021},
      eprint={2103.14769},
      archivePrefix={arXiv},
      primaryClass={q-fin.MF}
}

@article{blackscholes,
    author = {Fisher Black and Myron Scholes},
    title = {The pricing of options and corporate liabilities},
    journal = {Journal of political economy},
    year = {1973}
}

\end{document}